\documentclass{article}

\usepackage[preprint]{neurips_2019}
\usepackage[utf8]{inputenc} % allow utf-8 input
\usepackage[T1]{fontenc}    % use 8-bit T1 fonts
\usepackage{hyperref}       % hyperlinks
\usepackage{url}            % simple URL typesetting
\usepackage{booktabs}       % professional-quality tables
\usepackage{amsfonts}       % blackboard math symbols
\usepackage{nicefrac}       % compact symbols for 1/2, etc.
\usepackage{microtype}      % microtypography
\usepackage{natbib} 
\usepackage{caption} 
\usepackage{float}
\usepackage{graphicx}
\usepackage{subfig}
\usepackage{amsmath}

\title{Automated Audio Captioning and Language-Based Audio Retrieval}

\author{%
  Clive Gomes \\
  Carnegie Mellon University\\
  Pittsburgh, PA 15213 \\
  \texttt{cliveg@andrew.cmu.edu} \\
  \And  
  Hyejin Park \\
  Carnegie Mellon University\\
  Pittsburgh, PA 15213 \\
  \texttt{hyejinp@andrew.cmu.edu} \\
  \AND
  Patrick Kollman \\
  Carnegie Mellon University\\
  Pittsburgh, PA 15213 \\
  \texttt{pkollman@andrew.cmu.edu} \\ 
  \And
  Yi Song \\
  Carnegie Mellon University\\
  Pittsburgh, PA 15213 \\
  \texttt{yis2@andrew.cmu.edu} \\
  \AND
  Iffanice Houndayi \\
  Teaching Assistance\\
  \texttt{iffanice@cmu.edu} \\
  \And
  Ankit Shah\\
  Team Mentor\\
  \texttt{aps1@andrew.cmu.edu} \\
}
\begin{document}

\maketitle
    
\begin{abstract}
  This project involved participation in the DCASE 2022 Competition (Task 6) which had two subtasks: (1) Automated Audio Captioning and (2) Language-Based Audio Retrieval. The first subtask involved the generation of a textual description for audio samples, while the goal of the second was to find audio samples within a fixed dataset that match a given description. For both subtasks, the Clotho dataset was used. The models were evaluated on BLEU1, BLEU2, BLEU3, ROUGEL, METEOR, CIDEr, SPICE, and SPIDEr scores for audio captioning and R1, R5, R10 and mARP10 scores for audio retrieval. We have conducted a handful of experiments that modify the baseline models for these tasks. Our final architecture for Automated Audio Captioning is slightly better than the baseline performance, while our model for Language-Based Audio Retrieval has surpassed its counterpart.
  
\end{abstract}

\section{Introduction}

Automatic Audio Captioning (AAC) is defined as the problem of automatically generating a language description from sounds (\cite{drossos2017automated}); this description is called a caption. In this task, we aim to generate a caption as close as possible to human-perceived information for audio signals. Methods can model concepts such as loudness, physical descriptions of objects and environments, and intelligent descriptions such as frequency, e.g., “a large car honks three times as it passes an empty road”.
We also address language-based audio retrieval.
% The second task is language-based audio retrieval. 
Here, the goal is to find audio signals within a fixed dataset that match a given textual description. For both tasks, our goal is to develop robust models that can handle audio clips of varying lengths. 

The Clotho dataset will be used for training and evaluation of both tasks (\cite{drossos2020clotho}). For audio captioning, model outputs will be scored on the following metrics: BLEU1, BLEU2, BLEU3, ROUGEL, METEOR, CIDEr, SPICE, and SPIDEr; of these, the main metric is SPIDEr which is a linear combination of SPICE and CIDEr using a policy gradient method to optimize (\cite{liu2017improved}). The R1, R5, R10 and mARP10 metrics are used for audio retrieval.

\section{Literature Review}

\subsection{Audio Captioning}

Compared to video and image captioning, audio captioning has recently begun to receive attention in the area of intelligent audio research, particularly since the 2020 DCASE challenge (\cite{gebhard1automated}). In order to improve the performance of audio captioning models, several studies have used the seq2seq approach. In a recent paper by \cite{weck2021evaluating}, the authors looked into the use of off-the-shelf models in performing the audio-captioning task. They evaluated four embedding models (VGGish, YAMNet, OpenL3 and COALA) as encoders, three variations of embedding adapters (identity function, multi-layer perceptron and multi-head attention), four word embeddings (Word2Vec, GloVe, fastText and BERT), a transformer-based decoder, and their combinations in various settings. Results show that YAMNet performed best as an encoder and can be improved using a multi-head attention-based adapter. As for word embeddings, BERT provided the best results.

Among works that build an audio-captioning model from the ground up, key variations include choice of architecture for the encoder and decoder, various feature extraction techniques, using overlapping versus non-overlapping time segments and the use of regularization. The approach of \cite{wu2020audio} during the 2020 DCASE Challenge involved the use of a 10-layer CNN encoder with a Transformer as the decoder. \cite{xu2021sjtu} modified this architecture and used a single-layer GRU with temporal attention in place of the Transformer. More recent studies (\cite{mei2021audio}, \cite{labbeirit}) have attempted to use a fully recurrent encoder and decoder, owing to the limitation of CNNs in modeling long-ranged temporal information. Others have tried to use richer features to improve the performance of the audio-captioning system. For one, Tran et al. make use of both RNN and CNN blocks to capture temporal information as well as time-frequency patterns.

\cite{kim2019audiocaps} presents another encoder-decoder model, but one that includes semantic attention in the encoder phase. Using a large audio caption dataset, they extract words from the captions and apply the nearest neighbor approach to retrieve the nearest labels as attribute words. These attribute words are then included in the encoder phase as additional semantic information. There are various other ways to include semantic information in a model. \cite{eren2020audio} offer the idea that subject-verb embeddings can be extracted from the captions and then be included in the encoder phase as well. For both models, the additional semantic information was proven to enhance the performance of the model.  

In conjunction with semantic embeddings, \cite{eren2021audio} include another type of method sound event detection based audio captioning in their model. They apply pre-trained neural networks (PANNs) to generate PANN features and sound event vectors, which are concatenated as an input to the encoder. They use a combination of GRU and bidirectional GRU and for the encoder-decoder architecture. \cite{xu2021text} creates an Audio-Grounding dataset, which offers a correspondence series of \textit{audio – caption – sound event phrase – sound event timestamp segmentation}. Based on such, they propose the text-to-audio grounding task to classify and
localize particular sound events in an audio clip in a more cross-modal way.

\subsection{Language-Based Audio Retrieval}

The task of Audio Retrieval has received limited attention in literature. As such, only a handful of papers on this topic have been published. One such work is the study by \cite{koepke2022audio} which proposed two new benchmarks for text-based audio retrieval on the AudioCaps and Clotho datasets. These benchmarks were used to build baselines, and the benefits of multiple pre-trained audio expert networks were demonstrated. Realizing the limitation of keyword matching retrieval, \cite{song2020agricultural} focused on the inverted index of silence words, and designed a hybrid model with representation retrieval and semantic retrieval. Another interesting approach was proposed by \cite{oncescu2021audio} which involves learning cross-modal embeddings for the audio samples and text descriptions such that captions that describe the audio well would be closer to the audio sample in this shared embedding space. A more recent study \cite{xie2021unsupervised} proposed an unsupervised audio-text aligning framework 
between unaligned and annotated audio embeddings and word embeddings. The audio embedding is extracted from the audio clip using convolutional recurrent neural network (CRNN) 
and the word embedding extracted from the caption using Word2Vec. Event-phrase correspondences can be predicted by calculating the similarity of the clip caption pairs 
by averaging frame-word similarities.

\section{Model description}

\subsection{Audio Captioning}

Overall, our task is to accept an audio sample as input and output a suitable caption for that sample. The input can be denoted as $X \in \mathbb{R}^{T \times N_{features}}$ 
where $T$ is the number of frames and $N_{features}$ is the number of features. The output can be denoted as $Y \in \mathbb{R}^{I \times L_{words}}$ where $I$ is the number of captions and $L_{words}$ is the length of words.

We will use a Seq2Seq approach to model the relationships between the audio clips and text descriptions. 
Similar to \cite{ikawa2019neural}, we intend to use RNNs for the encoder-decoder framework where the encoder takes in features extracted from the audio, while the decoder converts it into a sequence of words. To improve the model performance, we will likely use pre-trained models to extract audio features, such as YAMNet. 

To train the model, we use cross-entropy loss where $y_{t}$is the ground truth word at time step t
$$
L_{CE}=-\frac{1}{T}\sum^{T}_{t=1}logp(y_{t}|y_{1:t-1},\theta)
$$

Model outputs will be scored on the following metrics: BLEU1, BLEU2, BLEU3, ROUGE$_{l}$, METEOR, CIDE$_{r}$, SPICE, and SPIDE$_{r}$ (\cite{mei2021audio}). BLEU$_{n}$ calculates the weighted geometric mean over different length n-grams. ROUGE$_{l}$ computes F-measures by counting the longest common subsequence. METEOR calculates the harmonic mean of precision and recall based on explicit word-to-word matches between the caption and given references.CIDE$_{r}$ measures the cosine similarity between term frequency and inverse document frequency. SPICE creates scene graphs for captions and calculates F-score based on tuples in the scene graphs. Proposed by \cite{liu2017improved}, SPIDE$_{r}$ combines the semantic stability of SPICE and the fluency of CIDEr, which enable easier optimization by using Monte Carlo rollouts instead of mixing MLE training. 

\subsection{Language-Based Audio Retrieval}

Audio retrieval involves retrieving audio signals with content that matches a given textual description. This task has two inputs: (1) a set of audio samples of varying lengths; and (2) a free-form text $X \in \mathbb{R}^{L_{words}}$, where $L_{words}$ is the length of the description. The output is a vector $Y \in \mathbb{R}^{N}$ containing a score for each audio sample (i.e., $N$ is the number of audio clips).

To perform language-based audio retrieval, we would use our model from the previous audio captioning task to first create captions for all audio files in our database. We would then use a similarity metric to score these descriptions based on the free-form input text. These scores would be used to rank the audio files and return the top-k results. 

%Then, the model is optimized with a triplet ranking loss (\cite{xie2021unsupervised}), where $\lambda$ is a margin hyper-parameter fixed to 1.
%L_{TP}=\frac{1}{K} \sum^{K}_{k=1}[\max(0, S(x_{k}, \hat{y}_{k}) - S(x_{k}, y_{k})+\lambda]\\
%+\max(0, S(\hat{x}, y_{k})-S(x_{k},y{k})+\lambda)]

We use the following evaluation metrics: R1 (recall score at top 1 retrieved result); R5 (recall score at top 5 retrieved result); R10 (recall score at top 10 retrieved result); mAP10 (mean average precision at top 10 retrieved results).

\section{Dataset} \label{sec:data}

The primary dataset for training and evaluation of both tasks is the Clotho dataset (\cite{drossos2020clotho}). This dataset contains captions for 6974 audio files (5 captions per audio); duration of these audios vary between 15 and 30 seconds while captions are 8 to 20 words long. These captions describe the events taking place in the audio (e.g., “a person is turning a map over and over”)—which is the desired output of our first task. The dataset has a development, validation, and evaluation split. 3840 audio files are used in the development split, 1046 audio files are used in the validation split, and 2088 are used in the evaluation split. The captions for each split can be found in a respective CSV file.

The audio samples are available as WAV, while the captions are available as CSV files. For use in the development of an audio captioning method, features have to be extracted from the WAV audio clips and the captions in the CSV files have to be pre-processed (e.g., punctuation removal). The extracted features and processed words are then matched to be input-output pairs.

Besides the Clotho dataset, we have also located a few other audio datasets, specifically for sound event detection (which we can use for our second approach to Audio Captioning). These include the Freesound Audio Tagging dataset[20] from Kaggle (80 categories of everyday sounds like “slam”, “squeak”, etc.), the Urban Sound 8K dataset[21] (10 classes of urban sounds like “dog bark”), and the Environmental Sounds dataset[22] (50 classes of sounds like 'rain', 'crickets', and so on). We may potentially find and use other datasets as we begin to go through research papers over the course of this project.

\section{Baseline Implementation}

\subsection{Audio Captioning}

The baseline model for Audio Captioning is a seq2seq transformer model consisting of 6 encoders and 6 decoder layers. The baseline repository can be found here: \url{https://github.com/felixgontier/dcase-2022-baseline}. The model begins with an audio adapter, which is an affine layer that is used to extract embeddings from the audio samples. This audio adapter takes in as input embeddings from a pre-trained VGGish model of dimension 128, and outputs embeddings of 768. The VGGish model is a feature embedding model for audio classification tasks, and the code for this model can be found here: \url{https://github.com/harritaylor/torchvggish/}. The transformer in the model is a Bart seq2seq transformer. The encoder of the transformer takes as an input these embeddings of dimension 768, and outputs a sequence of embeddings of the same length as the input sequence. Each transformer layer of the encoder outputs 768 features for each time-step of the input representation. The performance of the pre-trained weights for the baseline model can be seen in Table~\ref{table:pretrained}, and the performance of the baseline model trained from scratch can be seen in Table~\ref{table:baseline}.

\begin{table}[H]
\centering
% \makebox[0pt][c]{\parbox{1.2\textwidth}{%
    \begin{minipage}[b]{0.4\linewidth}
    \centering
    \caption{Pre-trained Baseline}
    \label{table:pretrained}
    \begin{tabular}{ll}
    \toprule
    %\multicolumn{2}{c}{Part}
    Metric & Score \\
    \midrule					
    BLEU$_1$ & 0.555 \\
    BLEU$_2$ & 0.358 \\
    BLEU$_3$ & 0.239 \\
    BLEU$_4$ & 0.156 \\
    METEOR$_1$ & 0.164 \\
    ROUGE$_L$ & 0.364 \\
    CIDE$_r$ & 0.358 \\
    SPICE$_1$ & 0.109  \\
    SPICE$_r$ & 0.233 \\
    \bottomrule
    \end{tabular}%
    \end{minipage}
    \begin{minipage}[b]{0.5\linewidth}
    \centering
    \caption{Baseline trained from Scratch}
    \label{table:baseline}
    \begin{tabular}{llll}
    \toprule
    %\multicolumn{2}{c}{Part} \\
    \cmidrule(r){1-3}
    Metric & Epoch 1 & Epoch 2 & Epoch 3 \\
    \midrule					
    BLEU$_1$ & 0.479 & 0.494 & 0.506 \\
    BLEU$_2$ & 0.272 & 0.322 & 0.331 \\
    BLEU$_3$ & 0.155 & 0.216 & 0.219 \\
    BLEU$_4$ & 0.069 & 0.132 & 0.135 \\
    METEOR$_1$ & 0.122 & 0.138 & 0.140 \\
    ROUGE$_L$ & 0.314 & 0.347 & 0.346 \\
    CIDE$_r$ & 0.103 & 0.212 & 0.237 \\
    SPICE$_1$ & 0.069 & 0.090 & 0.093 \\
    SPICE$_r$ & 0.086 & 0.151 & 0.165 \\
    \bottomrule
    \end{tabular}%
    \end{minipage}
% }}
\end{table}

% \begin{table}[h]
%   \caption{Pre-trained Baseline}
%   \label{table:pretrained}
%   \centering
%   \resizebox{80}{!}{%
%   \begin{tabular}{ll}
%     \toprule
%     %\multicolumn{2}{c}{Part} \\
%     \cmidrule(r){1-3}
%     Metric & Score \\
%     \midrule					
%     BLEU$_1$ & 0.555 \\
%     BLEU$_2$ & 0.358 \\
%     BLEU$_3$ & 0.239 \\
%     BLEU$_4$ & 0.156 \\
%     METEOR$_1$ & 0.164 \\
%     ROUGE$_L$ & 0.364 \\
%     CIDE$_r$ & 0.358 \\
%     SPICE$_1$ & 0.109  \\
%     SPICE$_r$ & 0.233 \\
%     \bottomrule
%   \end{tabular}%
%   }
% \end{table} 

% \begin{table}[h]
%   \caption{Baseline trained from Scratch}
%   \label{table:baseline}
%   \centering
%   \resizebox{130}{!}{%
%   \begin{tabular}{llll}
%     \toprule
%     %\multicolumn{2}{c}{Part} \\
%     \cmidrule(r){1-3}
%     Metric & Epoch 1 & Epoch 2 & Epoch 3 \\
%     \midrule					
%     BLEU$_1$ & 0.479 & 0.494 & 0.506 \\
%     BLEU$_2$ & 0.272 & 0.322 & 0.331 \\
%     BLEU$_3$ & 0.155 & 0.216 & 0.219 \\
%     BLEU$_4$ & 0.069 & 0.132 & 0.135 \\
%     METEOR$_1$ & 0.122 & 0.138 & 0.140 \\
%     ROUGE$_L$ & 0.314 & 0.347 & 0.346 \\
%     CIDE$_r$ & 0.103 & 0.212 & 0.237 \\
%     SPICE$_1$ & 0.069 & 0.090 & 0.093 \\
%     SPICE$_r$ & 0.086 & 0.151 & 0.165 \\
%     \bottomrule
%   \end{tabular}%
%   }
% \end{table}

The Clotho dataset is given as WAV and CSV files. Preprocesssing is required to extract the  VGGish embeddings from the Clotho dataset. This preprocessing code is also provided and can be found at \url{https://github.com/felixgontier/dcase-2022-preprocessing}

The decoder employs encoder outputs to generate the caption in an autoregressive manner. To do so, previously generated words are tokenized and transformed into embeddings as inputs to the decoder. In the baseline model, the tokenizer is pre-trained with a byte-pair-encoding process, i.e., each token corresponds to a sub-word from the model vocabulary instead of a full English word. This tokenizer has a vocabulary size of 50265 tokens. Each token in the past sequence is then associated to a feature vector through an embedding map, and input to the decoder. Each layer of the decoder attends on the previously generated tokens with self-attention, as well as on the entire encoder output sequence with cross-attention. The embedding dimension of each decoder layer is 768. Lastly, a classifier head consisting of one linear layer with softmax activation outputs a probability for each token of the vocabulary.

Caption evaluation is then performed using a version of the caption evaluation tools used for the MS COCO challenge. At evaluation, generation is performed through beam search, although greedy decoding is also provided in the code. These tools be found online at \url{https://github.com/audio-captioning/caption-evaluation-tools}.

\subsection{Audio Retrieval}

\begin{figure}[H]
    \centering
    \includegraphics[width=\linewidth]{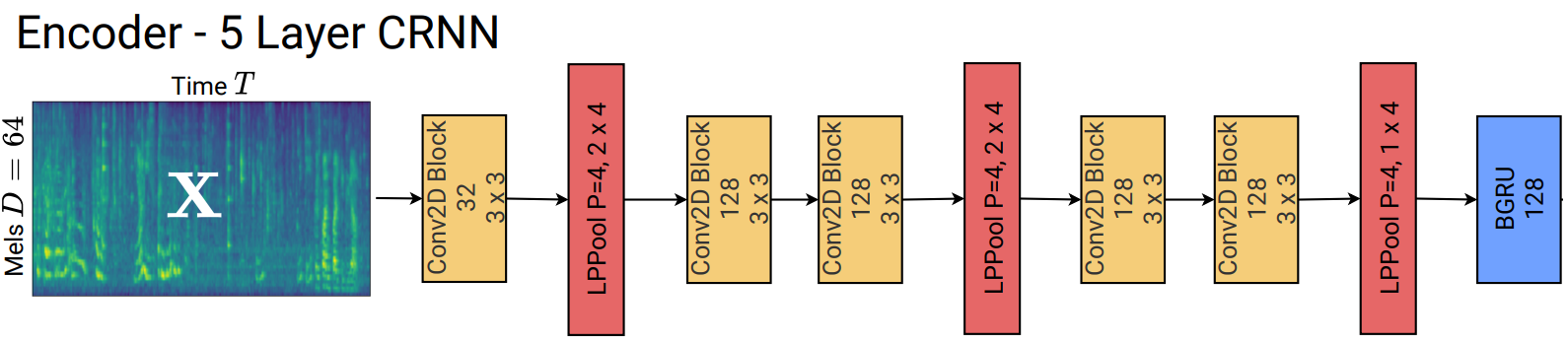}
    \caption{CRNN Encoder Architecture (\cite{xu2020crnn})}%
    \label{figure:crnn}
\end{figure}

The baseline model for this taskk involves an audio-to-text aligning framework (\cite{xie2021unsupervised}); a link to baseline repository is \url{https://github.com/xieh97/dcase2022-audio-retrieval}. The system use a CRNN encoder (a set of convolutional layers followed by one or more recurrent layers, as shown in Figure~\ref{figure:crnn}) to generate frame-level acoustic embeddings from audio clips (\cite{xu2020crnn}); 64 log mel features were used as the input which were computed on a sliding window of 40 ms and hop sizes of 20 ms. For the text (i.e., captions), a pre-trained Word2Vec model (trained on the Google News dataset) was used to transform textual descriptions into sequences of word embeddings \cite{mikolov2013efficient}. Both the acoustic embeddings and word embeddings were 300-long vectors and were averaged across time-steps to produce a single vector each. 

The idea behind this approach is to retrieve audio and text embedding having a high similarity score (dot product was used here) for matching pairs of audio and text descriptions and lower scores elsewhere. Triplet Margin loss was used to optimize the baseline, where the goal was to increase the margin between the similarity scores of matching pairs vs imposters (to select an imposter, a random audio or text was selected from within the training batch). 

The baseline model was trained using the Adam optimizer and a ReduceLROnPlateau schedulder for 50 epochs. Performance was measured in terms of recall at R1, R5, R10 and mAP10 (mean average precision). For the developement-evaluation split in Clotho dataset (\ref{sec:data}), the results of our baseline model can be seen in Table~\ref{table:retrival}. As can be seen, our baseline results were lower than what the original reference (*Add reference) listed. This could be due to various reasons: (1) the model was not trained long enough; (2) the hyperparameters used were not the best; (3) more regularization was needed to reduce overfitting; and so on. Nonetheless, we used the metrics obtained in our 50-epochs training as the baseline performance for the experiments we run.

\begin{table}[h] 
  \caption{Evaluation Metrics}
  \label{table:retrival}
  \centering
  \begin{tabular}{lll}
    \toprule
    %\multicolumn{2}{c}{Part} \\
    \cmidrule(r){1-3}
    Metric & Benchmark & Our Baseline \\
    \midrule					
    R1 & 0.03 & 0.02 \\
    R5 & 0.11 & 0.09 \\
    R10 & 0.19 & 0.16 \\
    mAP10 & 0.07 & 0.05 \\
    \bottomrule
  \end{tabular}
\end{table}

\section{Experiments}

\subsection{Audio Captioning}

Our basic workflow was as follows: we used the baseline model as a starting point, and in each experiment we made one or more modifications to the architecture. We compared the performance of our modifications to Table~\ref{table:baseline} to gauge whether the changes we made improvements to the baseline model performance. Given the limit of resources and GPU power, we halted the progress of training for an experiment if it wasn't comparable or better than the performance of the baseline model trained from scratch. Our code can be found here: \url{https://github.com/PatrickKollman/Deep-Learning-Final-Project}.

\subsubsection{BART vs CRNN Encoder}

The first modification we experimented with was choosing a Convolutional Recurrent Neural Network (CRNN) as our encoder instead of the Bart encoder. We chose to experiment with this encoder because it was the second-best encoder available to the team, and felt like a good starting point to understand the behavior of the model in relation to the data. We expected the results to vary because the data we are dealing with is time sensitive. The results seen in Table~\ref{table:crnn} confirm this speculation. The CRNN most likely struggled on the Clotho dataset because it is not passing time steps efficiently. The CRNN only contains one GRU encoder, which is much simpler compared to the complexity of the Bart transformer model.
\vspace{-1em}
\begin{table}[H]
  \caption{CRNN Encoder}
  \label{table:crnn}
  \centering
  \resizebox{100pt}{!}{%
  \begin{tabular}{ll}
    \toprule
    %\multicolumn{2}{c}{Part} \\
    \cmidrule(r){1-2}
    Metric & Epoch 1 \\
    \midrule					
    BLEU$_1$ & 0.318 \\
    BLEU$_2$ & 0.142 \\
    BLEU$_3$ & 0.065 \\
    BLEU$_4$ & 0.022 \\
    METEOR$_1$ & 0.100 \\
    ROUGE$_L$ & 0.253 \\
    CIDE$_r$ & 0.031 \\
    SPICE$_1$ & 0.035 \\
    SPICE$_r$ & 0.033 \\
    \bottomrule
  \end{tabular}%
  }
\end{table}

\subsubsection{Encoder Hidden State Output}

The second modification we examined was the intersection of the encoder and decoder. Instead of only passing the last hidden state of the encoder to the decoder, we took the mean of all of the hidden states of the encoder, and passed this average to the decoder. This modification was motivated by the concern that if the time sequences of our data were too long, a singular hidden state may be insufficient to capture all the information contained by the sequence. The results seen in Table~\ref{table:avg} argued otherwise. Since this modification yielded poor results, we hypothesized that the average of all the hidden states may have performed worse because the earlier hidden states contain little to no information. As a result, we continued by only averaging the last two hidden states of the encoder. The results seen in Table~\ref{table:two} show a minor improvement in performance to the baseline model.

\begin{table}[H]
\centering
% \makebox[0pt][c]{\parbox{1.2\textwidth}{%
    \begin{minipage}[b]{0.4\linewidth}
    \centering
    \caption{Hidden State Averaging}
    \label{table:avg}
    {\begin{tabular}{lll}
    \toprule
    %\multicolumn{2}{c}{Part} \\
    \cmidrule(r){1-3}
    Metric & Epoch 1 & Epoch 2 \\
    \midrule					
    BLEU$_1$ & 0.480 & 0.486 \\
    BLEU$_2$ & 0.260 & 0.315 \\
    BLEU$_3$ & 0.144 & 0.211 \\
    BLEU$_4$ & 0.058 & 0.127 \\
    METEOR$_1$ & 0.121 & 0.134 \\
    ROUGE$_L$ & 0.312 & 0.341 \\
    CIDE$_r$ & 0.087 & 0.198 \\
    SPICE$_1$ & 0.060 & 0.088 \\
    SPICE$_r$ & 0.074 & 0.143 \\
    \bottomrule
    \end{tabular}%
    }
    \end{minipage}
    \begin{minipage}[b]{0.5\linewidth}
    \centering
    \caption{2 Hidden States Averaged}
    \label{table:two}
    \begin{tabular}{llll}
    \toprule
    %\multicolumn{2}{c}{Part} \\
    \cmidrule(r){1-3}
    Metric & Epoch 1 & Epoch 2 & Epoch 3 \\
    \midrule					
    BLEU$_1$ & 0.474 & 0.497 & 0.508 \\
    BLEU$_2$ & 0.268 & 0.323 & 0.333 \\
    BLEU$_3$ & 0.153 & 0.217 & 0.221 \\
    BLEU$_4$ & 0.069 & 0.132 & 0.136 \\
    METEOR$_1$ & 0.119 & 0.138 & 0.143 \\
    ROUGE$_L$ & 0.313 & 0.348 & 0.349 \\
    CIDE$_r$ & 0.100 & 0.212 & 0.244 \\
    SPICE$_1$ & 0.067 & 0.090 & 0.094 \\
    SPICE$_r$ & 0.084 & 0.151 & 0.169 \\
    \bottomrule
    \end{tabular}%
    \end{minipage}
% }}
\end{table}

% \begin{table}[H]
%   \caption{Hidden State Averaging}
%   \label{table:avg}
%   \centering
%   \resizebox{100}{!}{%
%   \begin{tabular}{lll}
%     \toprule
%     %\multicolumn{2}{c}{Part} \\
%     \cmidrule(r){1-3}
%     Metric & Epoch 1 & Epoch 2 \\
%     \midrule					
%     BLEU$_1$ & 0.480 & 0.486 \\
%     BLEU$_2$ & 0.260 & 0.315 \\
%     BLEU$_3$ & 0.144 & 0.211 \\
%     BLEU$_4$ & 0.058 & 0.127 \\
%     METEOR$_1$ & 0.121 & 0.134 \\
%     ROUGE$_L$ & 0.312 & 0.341 \\
%     CIDE$_r$ & 0.087 & 0.198 \\
%     SPICE$_1$ & 0.060 & 0.088 \\
%     SPICE$_r$ & 0.074 & 0.143 \\
%     \bottomrule
%   \end{tabular}%
%   }
% \end{table}
% \begin{table}[H]
%   \caption{2 Hidden States Averaged}
%   \label{table:two}
%   \centering
%   \resizebox{130}{!}{%
%   \begin{tabular}{llll}
%     \toprule
%     %\multicolumn{2}{c}{Part} \\
%     \cmidrule(r){1-3}
%     Metric & Epoch 1 & Epoch 2 & Epoch 3 \\
%     \midrule					
%     BLEU$_1$ & 0.474 & 0.497 & 0.508 \\
%     BLEU$_2$ & 0.268 & 0.323 & 0.333 \\
%     BLEU$_3$ & 0.153 & 0.217 & 0.221 \\
%     BLEU$_4$ & 0.069 & 0.132 & 0.136 \\
%     METEOR$_1$ & 0.119 & 0.138 & 0.143 \\
%     ROUGE$_L$ & 0.313 & 0.348 & 0.349 \\
%     CIDE$_r$ & 0.100 & 0.212 & 0.244 \\
%     SPICE$_1$ & 0.067 & 0.090 & 0.094 \\
%     SPICE$_r$ & 0.084 & 0.151 & 0.169 \\
%     \bottomrule
%   \end{tabular}%
%   }
% \end{table}

\subsubsection{Increasing the Size of the Model}

We then explored the possibility that making the network deeper and more complex would improve performance. To do so, we ran one experiment which added a layer to the audio adapter and another separate experiment that increased the numbers of layers of encoders and decoders to 7. We initially expected this to not improve performance because we expected the creators of the baseline model to optimize these model hyper-parameters, but the results argued otherwise. Adding layers to both the audio adapter and transformer provided a significant increase in performance. These results can be seen in Table~\ref{table:adapter} and Table~\ref{table:transformer}.

\begin{table}[H]
\centering
% \makebox[0pt][c]{\parbox{1.2\textwidth}{%
    \begin{minipage}[b]{0.4\linewidth}
    \centering
    \caption{2 Layer Audio Adapter}
    \label{table:adapter}
    \resizebox{160pt}{!}
    {\begin{tabular}{llll}
    \toprule
    %\multicolumn{2}{c}{Part} \\
    \cmidrule(r){1-3}
    Metric & Epoch 1 & Epoch 2 & Epoch 3 \\
    \midrule					
    BLEU$_1$ & 0.497 & 0.501 & 0.509 \\
    BLEU$_2$ & 0.278 & 0.324 & 0.336 \\
    BLEU$_3$ & 0.155 & 0.216 & 0.225 \\
    BLEU$_4$ & 0.071 & 0.131 & 0.139 \\
    METEOR$_1$ & 0.124 & 0.136 & 0.141 \\
    ROUGE$_L$ & 0.322 & 0.343 & 0.348 \\
    CIDE$_r$ & 0.106 & 0.213 & 0.248 \\
    SPICE$_1$ & 0.070 & 0.091 & 0.093 \\
    SPICE$_r$ & 0.088 & 0.152 & 0.171 \\
    \bottomrule
    \end{tabular}%
    }
    % \hspace{3em}
    \end{minipage}
    \begin{minipage}[b]{0.5\linewidth}
    \centering
    \caption{7 Layer Encoder/Decoder}
    \label{table:transformer}
    \resizebox{160pt}{!}
    {\begin{tabular}{llll}
    \toprule
    %\multicolumn{2}{c}{Part} \\
    \cmidrule(r){1-3}
    Metric & Epoch 1 & Epoch 2 & Epoch 3 \\
    \midrule					
    BLEU$_1$ & 0.503 & 0.500 & 0.507 \\
    BLEU$_2$ & 0.280 & 0.325 & 0.337 \\
    BLEU$_3$ & 0.159 & 0.217 & 0.226 \\
    BLEU$_4$ & 0.070 & 0.132 & 0.141 \\
    METEOR$_1$ & 0.125 & 0.139 & 0.141 \\
    ROUGE$_L$ & 0.324 & 0.346 & 0.351 \\
    CIDE$_r$ & 0.118 & 0.220 & 0.248 \\
    SPICE$_1$ & 0.072 & 0.091 & 0.091 \\
    SPICE$_r$ & 0.095 & 0.156 & 0.170 \\
    \bottomrule
    \end{tabular}%
    }
    \end{minipage}
% }}
\end{table}

% \begin{table}[H]
%   \caption{2 Layer Audio Adapter}
%   \label{table:adapter}
%   \centering
%   \resizebox{130}{!}{%
%   \begin{tabular}{llll}
%     \toprule
%     %\multicolumn{2}{c}{Part} \\
%     \cmidrule(r){1-3}
%     Metric & Epoch 1 & Epoch 2 & Epoch 3 \\
%     \midrule					
%     BLEU$_1$ & 0.497 & 0.501 & 0.509 \\
%     BLEU$_2$ & 0.278 & 0.324 & 0.336 \\
%     BLEU$_3$ & 0.155 & 0.216 & 0.225 \\
%     BLEU$_4$ & 0.071 & 0.131 & 0.139 \\
%     METEOR$_1$ & 0.124 & 0.136 & 0.141 \\
%     ROUGE$_L$ & 0.322 & 0.343 & 0.348 \\
%     CIDE$_r$ & 0.106 & 0.213 & 0.248 \\
%     SPICE$_1$ & 0.070 & 0.091 & 0.093 \\
%     SPICE$_r$ & 0.088 & 0.152 & 0.171 \\
%     \bottomrule
%   \end{tabular}%
%   }
% \end{table}

% \begin{table}[H]
%   \caption{7 Layer Encoder/Decoder}
%   \label{table:transformer}
%   \centering
%   \resizebox{130}{!}{%
%   \begin{tabular}{llll}
%     \toprule
%     %\multicolumn{2}{c}{Part} \\
%     \cmidrule(r){1-3}
%     Metric & Epoch 1 & Epoch 2 & Epoch 3 \\
%     \midrule					
%     BLEU$_1$ & 0.503 & 0.500 & 0.507 \\
%     BLEU$_2$ & 0.280 & 0.325 & 0.337 \\
%     BLEU$_3$ & 0.159 & 0.217 & 0.226 \\
%     BLEU$_4$ & 0.070 & 0.132 & 0.141 \\
%     METEOR$_1$ & 0.125 & 0.139 & 0.141 \\
%     ROUGE$_L$ & 0.324 & 0.346 & 0.351 \\
%     CIDE$_r$ & 0.118 & 0.220 & 0.248 \\
%     SPICE$_1$ & 0.072 & 0.091 & 0.091 \\
%     SPICE$_r$ & 0.095 & 0.156 & 0.170 \\
%     \bottomrule
%   \end{tabular}%
%   }
% \end{table}

\subsubsection{Final Architecture}
\label{sec:final}

Pooling the conclusions of all the experiments, we chose a final architecture of 2 linear layers for the audio adapter, 7 layers for the encoder and decoder of the Bart model, and to average the last two hidden states of the encoder. We trained this architecture for 20 epochs and the results can be seen in Table~\ref{table:final}. Our training showed great success in the earlier epochs, producing scores that we're better than the scores of the baseline model trained from scratch. Unfortunately, this success plateaued in the later epochs, and our final architecture could not reach the performance of the baseline model with pre-trained weights. Even though our model's performance plateaued during training, we hypothesize that the model hasn't actually reached convergence yet. Given more time and resources, our team would train the model even further to find out if convergence had occurred or not.

\begin{table}[h]
  \caption{Final Architecture}
  \label{table:final}
  \centering
%   \resizebox{130}{!}
{%
  \begin{tabular}{lllll}
    \toprule
    %\multicolumn{2}{c}{Part} \\
    \cmidrule(r){1-5}
    Metric & Epoch 1 & Epoch 2 & Epoch 3 & Epoch 20  \\
    \midrule					
    BLEU$_1$ & 0.492 & 0.499 & 0.511 & 0.553  \\
    BLEU$_2$ & 0.283 & 0.325 & 0.336 & 0.357  \\
    BLEU$_3$ & 0.172 & 0.218 & 0.225 & 0.236 \\
    BLEU$_4$ & 0.086 & 0.131 & 0.139 & 0.147 \\
    METEOR$_1$ & 0.128 & 0.139 & 0.141& 0.159 \\
    ROUGE$_L$ & 0.330 & 0.343 & 0.349 & 0.361 \\
    CIDE$_r$ & 0.120 & 0.216 & 0.245 & 0.321 \\
    SPICE$_1$ & 0.070 & 0.094 & 0.92 & 0.104 \\
    SPICE$_r$ & 0.095 & 0.155 & 0.168 & 0.213 \\
    \bottomrule
  \end{tabular}%
  }
\end{table}

\subsection{Language-Based Audio Retrieval}

\begin{figure}[H]
    \centering
    \includegraphics[width=6.6cm, height=4cm]{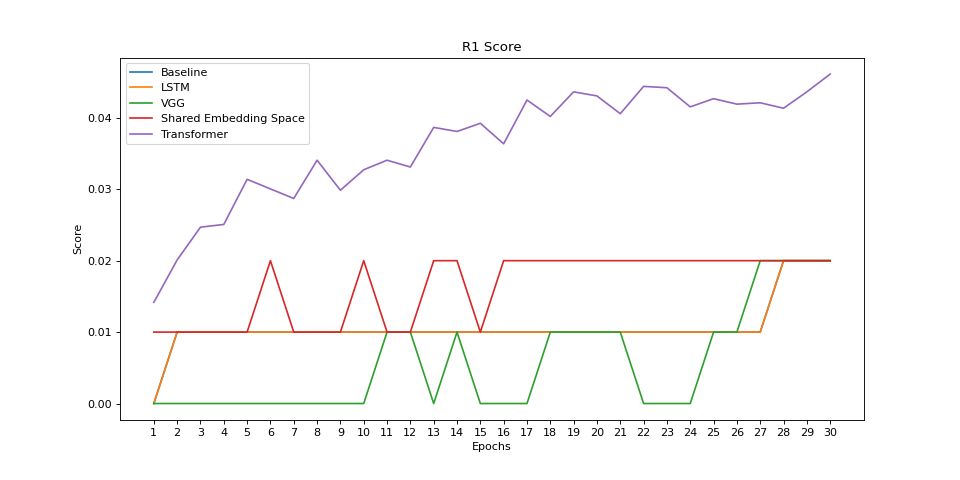}
    \includegraphics[width=6.6cm, height=4cm]{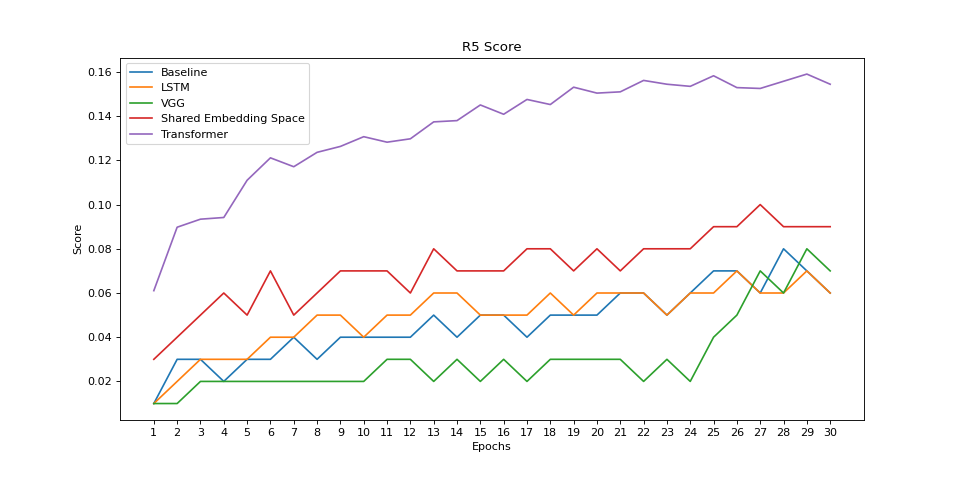}
    \includegraphics[width=6.6cm, height=4cm]{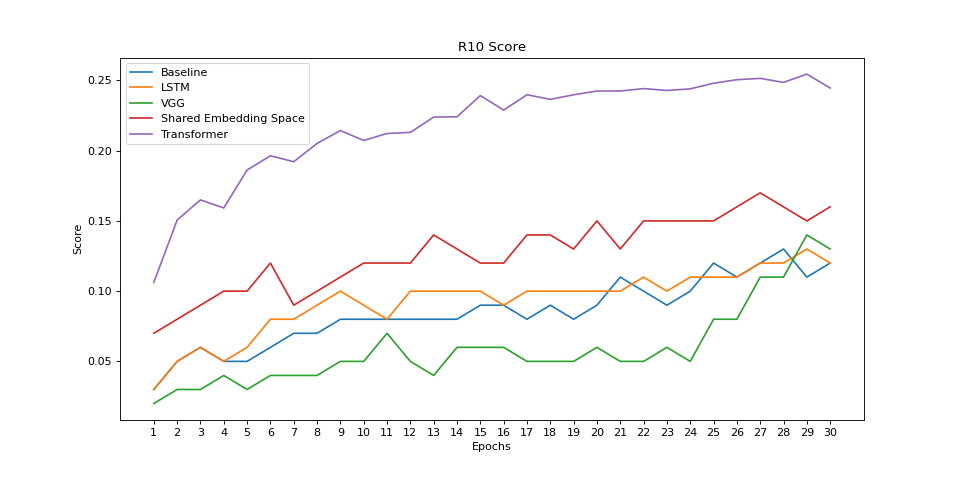}
    \includegraphics[width=6.6cm, height=4cm]{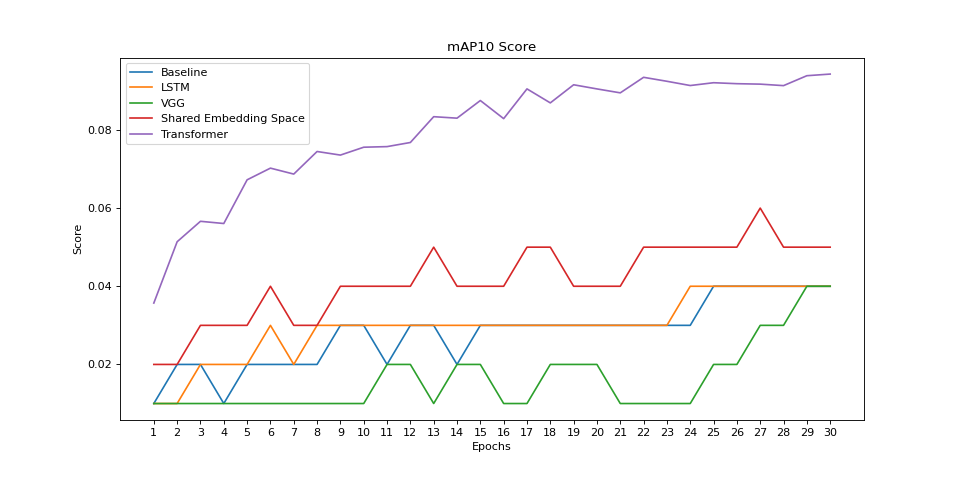}
    \caption{Evaluation Metrics for first 30 epochs (Audio Retrieval)}
    \label{figure:metrics}
\end{figure}

Similar to the Audio Captioning task, we conducted various experiments by modifying the baseline architecture and comparing our performance against the initial results. Each model was trained for up to 50 epochs, and the best scores were noted (i.e., we used early stopping based on validation scores). Results for all experiments have been summarized in Table~\ref{table:two}. Plots for the four metrics (R1, R5, R10 and mAP10) are also shown in Figure~\ref{figure:metrics}. Our code can be found here: \url{https://github.com/PatrickKollman/Deep-Learning-Final-Project}

\begin{table}[H]
  \caption{Evaluation Metrics for Task 2 Audio Retrieval}
  \label{table:two}
  \centering
%   \resizebox{300}{!}
{%
  \begin{tabular}{lllll}
    \toprule
    %\multicolumn{2}{c}{Part} \\
    \cmidrule(r){1-3}
   Model & R1 & R5 & R10 & mAP10 \\
    \midrule
    Baseline & 0.02 & 0.09 & 0.16 & 0.05 \\
    LSTM &  0.02 & 0.08 & 0.15 & 0.04 \\
    VGGish features & 0.02 & 0.08 & 0.14 & 0.04 \\
    Shared embedding space & \textbf{0.02} & \textbf{0.10} & \textbf{0.17} & \textbf{0.06} \\
    Sentence Transformer & \textbf{0.04}& \textbf{0.16} & \textbf{0.25} & \textbf{0.09} \\
    \bottomrule
  \end{tabular}%
  }
\end{table}

\subsubsection{GRU vs LSTM Cells}

First, we tried a very simple modification: the baseline CRNN network used a single GRU cell at the end of the CNN network; we simply replaced it with an LSTM cell and evaluated the resulting performance. The idea was that, since an LSTM cell has three rather than two gates, it could perform better than a GRU cell (at the cost of more training). However, we didn't observe any significant gains from using an LSTM (as seen in Table~\ref{table:two}). We believe this could be due to one of two reasons: (1) the baseline model is not complex enough to show an improvement (if any) from using an LSTM; or (2) the audio features used were encoded sufficiently well using a GRU and the added complexity of an LSTM was not required for this task. 

\subsubsection{Log-Mel vs VGGish features}

Apart from the standard log-mel features, we also tried using the VGGish features used in the Audio Captioning task as the input to our CRNN encoder. Since these features were specifically trained for audio tasks, we expected to obtain a better performance than when using log-mel features. Unfortunately, the model performed a bit poorly compared to the baseline. We believe this may be because the CRNN encoder is not complex enough to make proper use of the richer VGGish features and may need a transformer-like architecture, as in the Audio Captioning task, to get better results.

\subsubsection{Shared Embedding Space}
\label{sec:shared}

In the baseline implementation, we are using fixed text embeddings (i.e., Word2Vec) and learning embeddings for audio files that are close to those of text that describe them. The inherent issue here is that these embeddings for the audio and text, though close in value for similar pairs, are in different embedding spaces. Therefore, as proposed by \cite{elizalde2019cross}, we added an additional fully-connected layer at the end of our baseline model that takes the previous output text and audio embeddings, and learns new embeddings for both in a shared space (we converted the 300-long embedding to 1024 dimensions). As expected, we did indeed get an improvement, though not a lot. We believe that using a more complex architecture for the shared-embeddings model may lead to even better performance.

\subsubsection{Using Sentence Transformers for Text Embedding}
The Word2Vec embedding used in the previous experiments has shown great success as it can describe word similarities beyond simple syntactic regularities. However, there is one major limitation about Word2Vec vectors in that the representation of a single word is fixed once the training is done, which means that the word embedding is static. However, many words have difference meanings in difference contexts, which suggests that the word embedding should be dynamic and change according to the context. Transformers, owing to their self-attention mechanism, can generate word embedding that are fully contextualized. Popular transformer models such as BERT\cite{devlin2018bert} are often pre-trained on an enormously large amount of unstructured text data using two training tasks, sentence modeling(where a part of words are masked randomly and the transformer model is trained to predict the next word) and next sentence prediction. \\
\\
A large disadvantage of BERT is that it does not compute an independent sentence embedding, making it difficult to derive a good sentence level embedding from BERT-based models. A better method called S-BERT(sentence transformer) was proposed by \cite{reimers2019sentence} that uses Siamese network architecture to fine-tune a pre-trained BERT-like model with Natural Language Inference data. The sentence transformer has shown a great improvement over BERT-based transformer on modeling sentences. Two sentences with similar meaning will have similar embeddings.
% \\
We believe using embedding generated from the sentence transformer would yield a much better result in comparison to the Word2Vec vectors, as the embeddings will capture the meaning of the entire sentence. \\
\\
Since the competition allows us to use any pre-trained model, we used a pre-trained sentence transformer using the python sentence transformer library. In this approach, the text encoder part is not trainable (weights were frozen), and we used the triplet ranking loss to train the audio encoder, so it generates audio embedding that follows the distribution of the sentence embedding of the captions. As expected, the model performed significantly better than all previous experiments (as seen in Table~\ref{table:two}). The scores of R1, R5, R10, and mAP10 are improved by 50$\%$,77.8$\%$,56.3$\%$,80$\%$ respectively.

\subsubsection{Binary Cross Entropy Loss with Exponential Negative Euclidean Distance}
The baseline implementation uses a dot product as the similarity score between audio and text embeddings. \cite{elizalde2019cross} proposed a binary cross entropy loss with exponential euclidean distance as an alternative for the task of learning joint embeddings. The distance $d$ has the following equation for an audio text embedding pair $(a_i, t_i)$:

\begin{align*}
d = \exp\left(-\sqrt{\sum_{j=1}^{D} (a_{i,j}-t_{i,j})^2}\right) \;,
\end{align*}

where $D$ is the dimension of the embedding vectors. We think this loss could be better than the dot product because the magnitude of dot product is unbounded, while this score function computed using the exponential euclidean distance is bounded between 0 and 1. However, there wasn't sufficient time to test this in practice.

%\subsubsection{Sound Event Vectors}

\section{Conclusion}

Our experiments for both Audio Captioning and Language-Based Audio Retrieval yielded promising results. The \hyperref[sec:final]{final architecture} chosen for the Audio Captioning task (increased model size, and modified encoder output in the Bart transformer) yielded metrics that approached the baseline, while the \hyperref[sec:shared]{Shared-Embedding Space and Sentence Transformer} models have surpassed the performance of the baseline for Language-Based Audio Retrieval. 

In doing so, we have successfully addressed the problem of Automatic Audio Captioning and Language-Based Audio Retrieval. Our final model for the Automatic Audio Captioning task automatically generates a language description for sounds. As for Language-Based Audio Retrieval task, we have successfully ranked audio files based on textual descriptions, and the scores of R1, R5, R10, and mAP10 are improved by 50\%, 77.8\%, 56.3\%, 80\% respectively.

Given more time and resources, our team would have trained the final architecture for the Audio Captioning task even further to see if the plateau of scores was a false sign of convergence. We would also have pursued the use of sound event vectors as well. 
As in \cite{eren2021audio}, we extract sound event features using Cnn14\_DecisionLevelMax pretrained model. 
For Audio Captioning task, we made a vector of time step size and save the class of sound generated at each time step in this vector.
In addition, for Audio Retrieval task, we made a vector of the size of the number of classes, and whether each class occurred in the corresponding audio file was indicated in binary.
The findings from our research showed that including sound event vectors as an additional input to our models would provide valuable semantic information that would improve performance. Therefore, the performance of both tasks can be further improved by using the sound event vector.

\bibliographystyle{plainnat}
\bibliography{ref}

\end{document}